\documentclass[a4paper,10pt,twocolumn]{article}
\usepackage{amsmath}

\begin{document}

\title{\bf \Large 
Effective $N=1$ description of 
5D conformal supergravity\footnote{
Talk given by H.~Abe at Summer Institute '05, Yamanashi, 
Japan, 11-18 August, 2005}
}

\author{
Hiroyuki~Abe$^{1,}$\footnote{E-mail address: 
abe@gauge.scphys.kyoto-u.ac.jp} and 
Yutaka~Sakamura$^{2,}$\footnote{E-mail address: 
sakamura@het.phys.sci.osaka-u.ac.jp} \\*[10pt] 
$^1${\normalsize \it Department of Physics, Kyoto University, 
Kyoto 606-8502, Japan} \\
$^2${\normalsize \it Department of Physics, Osaka University, 
Osaka 560-0043, Japan}
}

\date{
\centerline{\small \bf Abstract}
\begin{minipage}{0.95\linewidth}
\medskip
\small
We construct an effective N=1 superfield description 
of five-dimensional conformal supergravity. 
By the use of this description, we reinterpret some 
physical systems such as Scherk-Schwarz supersymmetry 
breaking from the conformal supergravity viewpoint. 
We also show how to introduce a radion fluctuation mode 
in this framework. 
\end{minipage}
}

\maketitle 

The five-dimensional (5D) supergravity provides 
an interesting theoretical framework to the 
physics beyond the standard model (SM) in both 
bottom-up and top-down approach. 
In the former approach, the localized wavefunction 
can be the source of Planck and weak hierarchy~\cite{Randall:1999ee} 
and/or hierarchical structures within SM, 
the supersymmetry (SUSY) breaking sector 
can be hidden in the extra dimension, 
AdS/CFT correspondence provides a way to 
analyze perturbatively the four-dimensional (4D) 
strongly coupled theories which can be the extension 
of Higgs sector in SM, and so on. 
In the latter approach, it is known 
that in some case the compactified 
eleven-dimensional supergravity (M-theory) 
can be described effectively by the 5D supergravity 
at certain energy scale~\cite{Lukas:1998yy}. 
Also because the 5D supergravity is the simplest 
model with extra dimension, we have complete 
off-shell formulations (e.g., Ref.~\cite{Fujita:2001bd}) 
which allow a systematic study. 
In this talk we show an effective $N=1$ description 
of 5D conformal supergravity that will be useful 
in any approaches mentioned above.

First we briefly review the hypermultiplet 
compensator formulation of 5D conformal supergravity 
based on Ref.~\cite{Fujita:2001bd}. 
The 5D superconformal algebra consists of 
the Poincar\'e symmetry \mbox{\boldmath $P$}, \mbox{\boldmath $M$}, 
the dilatation symmetry \mbox{\boldmath $D$}, 
the $SU(2)$ symmetry \mbox{\boldmath $U$}, 
the special conformal boosts \mbox{\boldmath $K$}, 
$N=2$ supersymmetry \mbox{\boldmath $Q$}, 
and the conformal supersymmetry \mbox{\boldmath $S$}. 
The gauge fields corresponding to these generators 
$\mbox{\boldmath $X$}_{\!\!A}=\mbox{\boldmath $P$}_m$, 
$\mbox{\boldmath $M$}_{mn}$, 
$\mbox{\boldmath $D$}$, 
$\mbox{\boldmath $U$}_{ij}$, 
$\mbox{\boldmath $K$}_m$, 
$\mbox{\boldmath $Q$}_i$, 
$\mbox{\boldmath $S$}_i$, 
are respectively represented by 
$h_\mu^{\ A}=e_\mu^{\ m}$, 
$\omega_\mu^{mn}$, 
$b_\mu$, 
$V_\mu^{ij}$, 
$f_\mu^{\ m}$, 
$\psi_\mu^i$, 
$\phi_\mu^i$. 
We use $\mu,\nu,\ldots$ as five-dimensional curved indices 
and $m,n,\ldots$ as the tangent flat indices. 
The $i,j$ are $SU(2)_{\mbox{\boldmath \scriptsize $U$}}$ 
index and $\psi_\mu^i$ and $\phi_\mu^i$ are 
$SU(2)$ Majorana spinors. 
The relevant 5D superconformal multiplets 
to the following study are 
\begin{center}
\begin{tabular}{ll}
Weyl multiplet: & 
($e_\mu^{\ m}$, $\psi_\mu^i$, $V_\mu^{ij}$, 
$b_\mu$, $v^{mn}$, $\chi^i$, $D$), \\
Vector multiplet: & 
($M$, $W_\mu$, $\Omega^i$, $Y^{ij}$)$^I$, \\
Hypermultiplet: & 
(${\cal A}^\alpha_{\ i}$, $\zeta^\alpha$, 
${\cal F}^\alpha_{\ i}$), \\
\end{tabular}
\end{center}
where $I=0,1,2,\ldots,n_V$ and $\alpha=1,2,\ldots,2 (p+q)$. 
The $n_V+1$ is the number of vector multiplet and 
the $p$ ($q$) stands for the number of compensator 
(physical) hypermultiplet. 
The superconformal gauge fixing for a reduction to 
5D Poincar\'e supergravity is given by  
\begin{eqnarray}
\begin{array}{lcl}
\mbox{\boldmath $D$} &:& {\cal N} =M_5^3 \equiv 1, \\
\mbox{\boldmath $U$} &:& {\cal A}^a_{\ i} \propto \delta^a_{\ i}, 
\qquad (p=1) \\
\mbox{\boldmath $S$} &:& {\cal N}_I\Omega^{Ii}=0, \\
\mbox{\boldmath $K$} &:& {\cal N}^{-1}\hat{\cal D}_m {\cal N}=0, 
\end{array}
\label{eq:scgf}
\end{eqnarray}
where 
${\cal N}=C_{IJK} M^I M^J M^K$ 
is the norm function of 5D supergravity.

In the following we derive an effective $N=1$ superspace 
description of 5D conformal supergravity by considering 
4D Poincar\'e invariant background, 
$ds^2=e^{2\sigma(y)}\eta_{\underline\mu \underline\nu}
dx^{\underline\mu}dx^{\underline\nu}-dy^2$, 
$\psi^i_\mu=0$ and so on, and neglecting the fluctuations of 
all the 5D gravitational fields including the graviphoton. 
The invariant action is written in the $N=1$ superspace 
as~\cite{PaccettiCorreia:2004ri,Abe:2004ar} 
$S=\int d^5x\, ({\cal L}_V +{\cal L}_H +{\cal L}_{N=1})$ where 
\begin{eqnarray}
{\cal L}_V &=& 
\frac{3}{2} C_{IJK} \bigg[ 
\int d^2 \theta\, \Big\{ i\Phi_S^I {\cal W}^J {\cal W}^K 
\nonumber \\ &&
+\frac{1}{12}\bar{D}^2(V^I D^\alpha \partial_y V^J 
-D^\alpha V^I \partial_y V^J) {\cal W}^K_\alpha \Big\} 
\nonumber \\ &&
+\textrm{h.c.} \bigg] 
-e^{2\sigma}\int d^4 \theta\, V_T 
C_{IJK} {\cal V}_S^I {\cal V}_S^J {\cal V}_S^K, 
\nonumber
\end{eqnarray}
\begin{eqnarray}
{\cal L}_H &=& -2e^{2\sigma} \int d^4 \theta\, V_T 
d_\alpha^{\ \beta} \bar\Phi^\beta 
\Big( e^{-2igV^It_I} \Big)^\alpha_{\ \gamma} \Phi^\gamma 
\nonumber \\ &&
-e^{3\sigma} \bigg[ \int d^2 \theta\, \Phi^\alpha 
d_\alpha^{\ \beta} \rho_{\beta \gamma} 
\big( \partial_y-2g \Phi_S^I t_I \big)^\gamma_{\ \delta} 
\Phi^\delta 
\nonumber \\ &&
+\textrm{h.c.} \bigg], 
\nonumber
\end{eqnarray}
\begin{eqnarray}
{\cal L}_{N=1} &=& 
\sum_{l=0,\pi} \Lambda_l \delta(y-lR) \bigg[ 
\nonumber \\ &&
-\frac{3}{2} e^{2\sigma} 
\int d^4 \theta\, \bar\Sigma \Sigma e^{-K^{(l)}(S,\bar{S})/3} 
\nonumber \\ &&
+\bigg\{  
\int d^2 \theta \Big( f^{(l)}_{\bar{I}\bar{J}} 
{\cal W}^{\bar{I}}{\cal W}^{\bar{J}} 
+e^{3\sigma} \Sigma^3 W^{(l)}(S) \Big) 
\nonumber \\ &&
+\textrm{h.c.} \bigg\} \bigg]. 
\nonumber
\end{eqnarray}
The $N=1$ vector and chiral superfields $V^I$ and 
$\Phi_S^I$ come from the 5D vector multiplet and 
the $N=1$ chiral superfields $\Phi^\alpha$ originate 
from the 5D hypermultiplets. 
We can find the relation between the original 
superconformal multiplets and these superfields 
in Ref.~\cite{Abe:2004ar}. 
The compensator chiral multiplets are given by 
$\Sigma=(\Phi^{\alpha=2})^{2/3}$, 
$\Sigma^C=(\Phi^{\alpha=1})^{2/3}$, 
and we include spurious superfield, 
$V_T=
\langle e_y^{\ 4} \rangle 
+i\theta^2 e^{\sigma} 
\langle V^{(1)}_y +iV^{(2)}_y \rangle 
-i\bar\theta^2 e^{\sigma} 
\langle V^{(1)}_y -iV^{(2)}_y \rangle$. 
The superfield $S$ symbolically represents 
boundary (induced or own) chiral superfields. 
From $V^I$ and $\Phi_S^I$, we can construct 
gauge invariant superfields, 
${\cal W}^I_\alpha =-\frac{1}{4}\bar{D}^2 D_\alpha V^I$, 
${\cal V}_S^I=V_T^{-1} 
\big( -\partial_y V^I -i(\Phi_S^I-\bar\Phi_S^I) \big)$. 
Note that the gauge groups are limited to the 
Abelian group in the above action for simplicity. 

Next we derive the action after the 
superconformal gauge fixing (\ref{eq:scgf}). 
We assume the standard form of ${\cal N}$, 
\begin{eqnarray}
{\cal N} &=& (M^{I=0})^3 -\frac{1}{2}M^{I=0} 
\sum_{x=1}^{n_V-1}(M^{I=x})^2, 
\nonumber
\end{eqnarray}
and general hypermultiplet gaugings, 
\begin{eqnarray}
\big(\,T_{I=0},T_{I=x}\,\big)\varphi^2 &=&
\big({\textstyle{-\frac{3}{2}}k}\,\epsilon(y),\, 
-r_x \big)i\sigma_3 \varphi^2, 
\nonumber \\
\big(\,T_{I=0},T_{I=x}\,\big)\varphi^{2v+2} &=& 
\big(c\,\epsilon(y),\, q_x\big)i\sigma_3 
\varphi^{2v+2}. 
\nonumber
\end{eqnarray}
Then at the leading order in an expansion 
in powers of $1/M_5$, the vector field in $V^{I=0}$ 
becomes the graviphoton that we neglect the 
fluctuation, and we find 
$V^{I=0} \simeq 0$, 
$\Phi_S^{I=0} \simeq {\cal T}$, 
$V_T \simeq \frac{{\cal T}+\bar{{\cal T}}}{2}$, 
$\Sigma \simeq 1-\theta^2 {\cal F}_\Sigma$ and 
$\Sigma^C \simeq -\theta^2 {\cal F}^C_\Sigma$, 
where 
${\cal T}=1-\theta^2{\cal F}_T$ and 
${\cal F}_T=-2i\langle V_y^{(1)}+iV_y^{(2)} \rangle$. 
The action after the gauge fixing~\cite{Abe:2004ar} 
is found to be 
$S=\int d^5x\, 
({\cal L}_{V'+H'} 
+{\cal L}_{N=1} 
+{\cal L}_{V_0+H_0} 
+{\cal L}_{\rm SB})$ where 
\begin{eqnarray}
{\cal L}_{V'+H'} &=& 
\left\{ \int d^2\theta\, \frac{1}{4} 
{\cal T} {\cal W}^x {\cal W}^x +\textrm{h.c.} \right\} 
\nonumber \\ &&
+e^{2\sigma} \int d^4 \theta\, 
\frac{2}{{\cal T}+\bar{{\cal T}}} 
\left( \partial_y V^x 
-\frac{\chi^x +\bar\chi^x}{\sqrt{2}} \right)^2 
\nonumber \\ &&
+e^{2\sigma}\int d^4 \theta\, 
\frac{{\cal T}+\bar{{\cal T}}}{2} 
\Big( \bar{H}^v e^{2q_x V^x} H^v 
\nonumber \\ &&
+\bar{H}^{Cv} e^{-2q_x V^x} H^{Cv} \Big) 
\nonumber \\ &&
+e^{3\sigma} \bigg\{ 
\int d^2 \theta\, H^{Cv} 
\Big( \frac{1}{2} \overleftrightarrow{\partial_y} 
+c\,{\cal T} 
\nonumber \\ &&
+\sqrt{2}q_x \chi^x \Big) H^v 
+\textrm{h.c.} \bigg\} 
\nonumber
\end{eqnarray}
\begin{eqnarray}
{\cal L}_{N=1} &=& 
\sum_{l=0,\pi}\Lambda_l \delta(y-lR) \Bigg[
\nonumber \\ &&
-\frac{3}{2} e^{2\sigma}\int d^4\theta\, 
(\bar\Sigma e^{\frac{4}{3}r_x V^x} \Sigma \, 
e^{-\frac{1}{3}K^{(l)}(S,\bar{S})} 
\nonumber \\ &&
+\bigg\{ \int d^2\theta\, 
\Big( f^{(l)}_{\bar{x}\bar{y}}(S) 
{\cal W}^{\bar{x}} {\cal W}^{\bar{y}} 
\nonumber \\ &&
+e^{3\sigma} \Sigma^3 W^{(l)}(S) \Big) 
+\textrm{h.c.} \bigg\} \Bigg], 
\nonumber
\end{eqnarray}
\begin{eqnarray}
{\cal L}_{V_0+H_0} &=& 
-2 e^{2\sigma}\int d^4 \theta\, 
\frac{{\cal T}+\bar{{\cal T}}}{2} 
\Big( \bar\Sigma^{\frac{3}{2}} 
e^{2r_x V^x} \Sigma^{\frac{3}{2}} 
\nonumber \\ &&
+(\bar\Sigma^C)^{\frac{3}{2}} 
e^{-2r_x V^x} (\Sigma^C)^{\frac{3}{2}} \Big) 
\nonumber \\ &&
-2e^{3\sigma} \bigg\{ 
\int d^2 \theta\, (\Sigma^C)^{\frac{3}{2}} 
\Big( \frac{1}{2} \overleftrightarrow{\partial_y} 
-\frac{3}{2}k\,{\cal T} 
\nonumber \\ &&
+\sqrt{2}r_x \chi^x \Big) \Sigma^{\frac{3}{2}} 
+\textrm{h.c.} \bigg\} 
\nonumber \\ &&
-8e^{2\sigma} \int d^4 \theta\, 
\frac{{\cal T}+\bar{{\cal T}}}{2}, 
\label{eq:gfaction}
\end{eqnarray}
where 
$H^v=\sqrt{2}\Phi^{\alpha=2v+2}$, 
$H^{Cv}=\sqrt{2}\Phi^{\alpha=2v+1}$, 
$V^x=V^{I=x}/\sqrt{2}$ and 
$\chi^x=-i\Phi_S^{I=x}$. 
The last term in the action 
can not be written in terms of 
the $N=1$ superfields and is given by~\cite{Abe:2004ar} 
\begin{eqnarray}
{\cal L}_{\rm SB} &=& 
e^{4\sigma} f_G \Big\{ 
(\partial_y +3\dot\sigma+3k-f_G)(M^x)^2 
\nonumber \\ && \qquad 
+\frac{3}{2} \big( \partial_y
+\frac{5}{2} k 
+\frac{3}{2} \dot\sigma 
\big) (|h|^2+|h^C|^2) 
\nonumber \\ && \qquad 
-(\sqrt{2}q_x M^x-c)(|h|^2-|h^C|^2) 
\nonumber \\ && \qquad 
+{\textstyle \frac{e^{-2\sigma}}{4}} 
(\chi_S^x \lambda^x +\textrm{h.c.}) \Big\}, 
\nonumber
\end{eqnarray}
where 
$f_G=\dot\sigma-\frac{2}{3} 
\langle {\cal N}_I Y^{I(3)} \rangle$. 
For BPS background, $f_G$ vanishes 
and then ${\cal L}_{\rm SB}=0$. 
Because $f_G$ represents the deviation 
of the background geometry from the 
BPS one, we conclude that ${\cal L}_{\rm SB}$ 
describes the effect of the geometry 
mediated SUSY breaking. 

Another immediate result from the action 
(\ref{eq:gfaction}) is about the 
Scherk-Schwarz (SS) SUSY breaking~\cite{next}. 
SS SUSY breaking is the consequence 
of the twisted boundary condition 
$\Phi(x,y+2\pi R)
=e^{-i\pi \vec\omega \cdot \vec\sigma}\Phi(x,y)$ 
where $\vec\omega=(\omega_1,\omega_2,\omega_3)$ 
is the twist vector and the Pauli matrices 
$\vec\sigma=(\sigma_1,\sigma_2,\sigma_3)$ 
acts on the $SU(2)_R$ index of the field $\Phi$ 
in the Poincar\'e supergravity. 
In the conformal supergravity point of view, 
the $SU(2)_R$ symmetry is the diagonal subgroup of 
$SU(2)_{\mbox{\boldmath \scriptsize $U$}} \times SU(2)_\Sigma$ 
determined by the $\mbox{\boldmath $U$}$-gauge fixing 
condition in (\ref{eq:scgf}), where $SU(2)_\Sigma$ 
is the rotation in terms of $a$ index in the 
compensator hypermultiplet ${{\cal A}^a_{\ i}}$. 
Then the above SS twist is physically equivalent 
to the twisted $\mbox{\boldmath $U$}$-gauge fixing 
${\cal A}^a_{\ i} \propto \delta^a_{\ i} 
\to (e^{i \pi \vec\omega \cdot \vec\sigma \alpha(y)})^a_{\ i}$ 
where $\alpha(y)$ is a gauge fixing parameter which 
satisfies $\alpha(y+2\pi R)=\alpha(y)+2\pi$. 
The relevant part to this change in the action 
(\ref{eq:gfaction}) is the $y$-derivative term of 
$\Sigma$, $\Sigma^C$ and we find an additional term like 
$\partial_y \alpha(y) \left\{ 
\int d^2 \theta\, \Sigma^3 W_0 
+\textrm{h.c.} \right\}$, 
where $W_0=\omega_1+i\omega_2$. 
Here we can choose $\alpha(y)$ in two 
distinctive ways. For $\alpha(y)=y/R$, 
the additional term is the bulk constant 
superpotential, while $\alpha(y)=\frac{1}{2}\pi 
\sum_n ({\rm sgn}(y-n \pi R)
-{\rm sgn}(-n \pi R))$ results in 
the constant superpotential at 
the boundaries because 
$\partial_y \alpha(y) 
=\pi \sum_n \delta(y-n \pi R)$. 
Then in the superconformal framework, 
we find directly that the SS SUSY 
breaking is equivalent to the 
constant superptential. 
Note that the compensator must have vanishing 
gauge charges $k,r_x=0$ for the case with 
the twisted $\mbox{\boldmath $U$}$-gauge fixing, 
otherwise the corresponding gauge field 
acquires a nonvanishing mass without 
the Higgs mechanism. Namely the SS twist 
basically conflicts with AdS$_5$ 
geometry~\cite{Hall:2003yc}.

We finally consider how to include the 
radion fluctuation mode in the previous 
$N=1$ superspace action. 
We start from the metric $ds^2=e^{2F(b(x),y)} 
\eta_{\underline\mu \underline\nu}
dx^{\underline\mu \underline\nu}
-G^2(b(x),y)dy^2$ with the radion 
fluctuation mode $b(x)$ by assuming 
a BPS radion stabilization mechanism 
(e.g., Ref.~\cite{Maru:2003mq}) with 
a small backreaction. 
The background geometry of our system 
is AdS, and the function $F$ and $G$ 
should satisfy 
$F(\langle b \rangle,y)=\sigma(y)=-ky$ 
and 
$G(\langle b \rangle,y)=1$. 
Because $b(x)$ is a modulus field, 
we also require $\partial_y F=-kG$. 
The embedding of $b(x)$ into the 
$N=1$ superspace action is done by 
the replacement $\sigma(y) \to F(b(x),y)$ 
and $\langle e_y^{\ 4} \rangle \to G(b(x),y)$ 
in the previous results. This yields a 
kinetic term for $b(x)$ after the 
superconformal gauge fixing 
which is compared to the corresponding one 
in the original supergravity action. 
The matching condition of these two 
as well as the above AdS and modulus 
conditions determines the 
$b(x)$-dependence of $F$ and $G$ as 
$F=\frac{1}{2}\ln (e^{2\sigma(y)}+b(x))$, 
$G=(1+e^{-2\sigma(y)}b(x))^{-1}$. 
The radion field itself should 
correspond to the proper length of 
the extra dimension, i.e., 
$r(x)=\frac{1}{\pi} \int_0^{\pi R}dy\,G$. 
Then the relation between $r(x)$ and $b(x)$ 
is given by 
$b(x)=e^{-k\pi R} \frac{\sinh \pi k (R-r(x))}
{\sinh \pi k r(x)}$. 

By promoting the radion field $r(x)$ to the 
superfield $T(x)$, we obtain the $N=1$ superspace 
action with the dynamical radion superfield: 
\begin{eqnarray}
{\cal L}_{V'+H'} &=& 
\left\{ \int d^2\theta\, \frac{1}{4} 
G(T) {\cal W}^x {\cal W}^x +\textrm{h.c.} \right\} 
\nonumber \\ &&
+e^{2\sigma} \int d^4 \theta\, 
G_R^{-2}(T) \left( \partial_y V^x 
-\frac{\chi^x +\bar\chi^x}{\sqrt{2}} \right)^2 
\nonumber \\ &&
+e^{2\sigma}\int d^4 \theta\, 
G_R^{3/2}(T) \Big( \bar{H}^v e^{2q_x V^x} H^v 
\nonumber \\ &&
+\bar{H}^{Cv} e^{-2q_x V^x} H^{Cv} \Big) 
\nonumber \\ &&
+e^{3\sigma} \bigg\{ 
\int d^2 \theta\, H^{Cv} 
\Big( \frac{1}{2} 
\overleftrightarrow{\partial_y} +c\,G(T) 
\nonumber \\ &&
+\sqrt{2}q_x \chi^x \Big) H^v 
+\textrm{h.c.} \bigg\}, 
\nonumber
\end{eqnarray}
\begin{eqnarray}
{\cal L}_{N=1} &=& 
\sum_{l=0,\pi}\Lambda_l \delta(y-lR) \bigg[
\nonumber \\ &&
-\frac{3}{2} e^{2\sigma}\int d^4\theta\, 
G_R^{-1}(T) e^{-\frac{1}{3}K^{(l)}(S,\bar{S})} 
\nonumber \\ &&
+\bigg\{ \int d^2\theta\, \Big( 
f^{(l)}_{\bar{x}\bar{y}}(S) 
{\cal W}^{\bar{x}} {\cal W}^{\bar{y}} 
\nonumber \\ &&
+e^{3\sigma} G^{-\frac{3}{2}}(T) 
W^{(l)}(S) \Big) +\textrm{h.c.} 
\bigg\} \bigg], 
\nonumber
\end{eqnarray}
\begin{eqnarray}
{\cal L}_{V_0+H_0} &=& 
-3e^{2\sigma}\int d^4 \theta\, 
\ln G_R(T), 
\nonumber
\end{eqnarray}
where $G_R(T)=\frac{G(T)+G(\bar{T})}{2}$. 
Based on this action, we can calculate 
the radion mass for the radion stabilization 
mechanism proposed by Ref.~\cite{Maru:2003mq}. 
We introduce a superpotential 
$W^{(l)}(S)=J_lS$ at the orbifold 
fixed points $y=lR$ ($l=0,\pi$) where  
$S=G^{\frac{3}{4}}(T)H$ and 
$H$ is a stabilizer hypermultiplet. 
We can easily obtain the 4D effective 
action in superspace for the zero mode 
$h_{(0)}$ of bulk hypermultiplet $H$ 
and the radion $T$, and then find a BPS 
vacuum $\langle h_{(0)} \rangle =0$ and 
$J_0-J_\pi e^{-(\frac{3}{2}k+c)\pi R}=0$ 
where $\langle T \rangle=R$. 
The radion mass on this BPS vacuum is 
derived up to ${\cal O}(|J_\pi|^2)$ as 
$m_{\rm rad}^2 
=\frac{k^2|J_\pi|^2}{6} 
\left( 1-\frac{2c}{k} \right) 
\left( \frac{3}{2}-\frac{c}{k} \right)^2 e^{-2k \pi R} 
\frac{1-e^{-2k\pi R}}{1-e^{-(k-2c)\pi R}}$. 
We find that the radion mass is finite even 
in $k \to 0$ limit, namely the stabilization 
mechanism can work in the flat spacetime.

In summary, we have derived an effective $N=1$ 
description of 5D conformal supergravity, 
and analyzed some SUSY breaking configurations 
such as geometry-mediated or Scherk-Schwarz 
breaking in this framework. 
We have also shown how to include dynamical 
radion mode in the $N=1$ superspace action. 
This result will be useful for the phenomenological 
studies of 5D supergravity as well as the 
theoretical understanding~\cite{next}. 
An application to the 5D supergravity with parity odd 
couplings such as boundary FI terms~\cite{Abe:2004nx} 
and the Green-Schwarz mechanism~\cite{Dudas:2004ni} 
would be fruitful. 

\section*{Acknowledgments}
This work was supported by KRF PBRG 2002-070-C00022 (H.A.), 
and the Astrophysical Research Center for the Structure 
and Evolution of the Cosmos (ARCSEC) funded by the Korea 
Science and Engineering Foundation and the Korean Ministry 
of Science (Y.S.).

\end{document}